\documentclass[twocolumn,superscriptaddress,preprintnumbers,nofootinbib,amsmath,amssymb]{revtex4}

\usepackage{graphicx}
\usepackage{bm}
\usepackage{epsfig}

\begin{document}

\title{\boldmath
  Probing Heavy-Quark Spin Symmetry in Double $J/\psi$ Hadroproduction
  \unboldmath}

\author{Zhi-Guo He}
\affiliation{Department of Physics and Electronics, School 
	of Mathematics and Physics, Beijing University of 
	Chemical Technology, Beijing 100029, China}
\affiliation{{II.} Institut f\"ur Theoretische Physik, 
	Universit\"at Hamburg,
	Luruper Chaussee 149, 22761 Hamburg, Germany}
\author{Xiao-Bo Jin}
\affiliation{Department of Physics and Electronics, School of Mathematics 
	and Physics, Beijing University of Chemical Technology, 
	Beijing 100029, China}
\author{Bernd A. Kniehl}
\affiliation{{II.} Institut f\"ur Theoretische Physik, Universit\"at Hamburg,
	Luruper Chaussee 149, 22761 Hamburg, Germany}

\date{\today}

\begin{abstract}
We perform the first complete $\mathcal{O}(\alpha_s^5)$ analysis of the prompt hadroproduction of $J/\psi$ pairs with large transverse momenta $p_T^{\psi\psi}$ in the nonrelativistic-QCD factorization framework, including all possible Fock state combinations $c\bar{c}(m)+c\bar{c}(n)$, with $m,n={}^3S_1^{[1,8]},{}^1S_0^{[8]},{}^3P_J^{[1,8]}$.
We observe that CMS and ATLAS data constrain a specific linear combination of the long-distance matrix elements (LDMEs) $\langle\mathcal{O}^{J/\psi}({}^1S_0^{[8]})\rangle$ and $\langle\mathcal{O}^{J/\psi}({}^3P_0^{[8]})\rangle$.
In conjunction with two other combinations fixed by single prompt $J/\psi$ hadroproduction, we gain a new LDME set, which turns out to be largely compatible with the world data of prompt $J/\psi$ yield and polarization and to probe heavy-quark spin symmetry, by which agreement is established with LHCb data of prompt $\eta_c$ yield.
Our $\mathcal{O}(\alpha_s^5)$ predictions also nicely agree with CMS and ATLAS data in the lowest bins of $J/\psi$ pair invariant mass $m^{\psi\psi}$, beyond leading-order kinematics.

\end{abstract}

\maketitle

More than half a century after the discovery of $J/\psi$ meson, the production 
mechanism of heavy quarkonium ($H$) remains mysterious.
The non-relativistic QCD (NRQCD) factorization formalism~\cite{Bodwin:1994jh} constructed on top of NRQCD effective field theory~\cite{Caswell:1985ui} provides an elegant and rigorous framework to separate perturbative and non-perturbative effects in heavy-quarkonium production.
Each possible $Q\bar{Q}$ Fock state $n={}^{2S+1}\!L_{J}^{[a]}$, in color singlet (CS) or color octet (CO) configuration, with $a=1,8$, contributes to the $H$ production cross section through a product of a process dependent short-distance coefficient (SDC), which comes as a perturbative series in the strong-coupling constant $\alpha_s$, and a supposedly universal non-perturbative long-distance matrix element (LDME), which is weighed by powers of the relative $Q$ quark velocity, with $v^2\approx0.3$ for charmonium, according to scaling rules \cite{Lepage:1992tx}.
Prompt $J/\psi$ production comprises direct production, through the hadronization of $c\bar{c}$ pairs, and feed-down, through decays of the $\chi_{cJ}$ and $\psi^\prime$ charmonia being directly produced on their own. 
The leading Fock states include $n={}^3\!S_1^{[1,8]},{}^1\!S_0^{[8]},{}^3\!P_J^{[8]}$ for $H=J/\psi,\psi^\prime$ and $n={}^3\!P_J^{[1]},{}^3\!S_1^{[8]}$ for
$H=\chi_{cJ}$ \cite{Bodwin:1994jh}.
Depending on production mode and phase space region considered, CO processes may well dominate, which has led to the term CO mechanism (COM).
The most prominent such example is arguably $J/\psi+W$ associated hadroproduction, where CS channels only contribute beyond next-to-leading order (NLO) in $\alpha_s$, while the NLO prediction is compatible with experimental data \cite{Butenschoen:2022wld}.

The verification of the LDME universality hypothesis has enjoyed top priority on the agenda of the world-wide heavy-quarkonium community for decades, with experimentalists and theorists joining forces \cite{QuarkoniumWorkingGroup:2004kpm,Lansberg:2006dh,Brambilla:2010cs,Butenschoen:2012qr,Brambilla:2014jmp,Lansberg:2019adr}.
So far, $J/\psi$ LDMEs have been determined through NLO fits to data of single, mostly prompt, production in $p\bar{p}$, $pp$, $\gamma p$, $\gamma\gamma$, and $e^+e^-$ collisions \cite{Ma:2010yw,Butenschoen:2010rq,Butenschoen:2011yh,Chao:2012iv,Gong:2012ug,Bodwin:2014gia,Shao:2014yta}.

Based on the results of Refs.~\cite{Butenschoen:2022wld,Butenschoen:2010rq,Butenschoen:2011yh,Butenschoen:2009zy,Butenschoen:2011ks,Butenschoen:2012px,Butenschoen:2014dra,Butenschoen:2019lef,Butenschoen:2020mzi}, such a fit was recently performed \cite{Brambilla:2024iqg} to CMS data of prompt $J/\psi$ yield at large transverse momentum $p_T^{J/\psi}$ \cite{CMS:2015lbl} and LHCb data on prompt $\eta_c$ yield \cite{LHCb:2014oii,LHCb:2019zaj} leading to good agreement with the $J/\psi$ world data, with the exception of low- and medium-$p_T^{J/\psi}$ hadroproduction, photoproduction at inelasticity $z > 0.6$, and large-$p_T^{J/\psi}$ $J/\psi+Z$ associated hadroproduction.
This comes as silver linings on the horizon in the as-yet confusing overall picture of LDME universality.
In Ref.~\cite{Brambilla:2024iqg}, heavy-quark spin symmetry (HQSS) was assumed to relate $J/\psi$ and $\eta_c$ LDMEs.
The impact of the LHCb data \cite{LHCb:2014oii} on the $J/\psi$ LDMEs was originally emphasized in Refs.~\cite{Butenschoen:2014dra,Han:2014jya,Zhang:2014ybe}.

In this work, we discover that the good overall description of $J/\psi$ data \cite{Brambilla:2024iqg} can be maintained without recourse to HQSS, by injecting orthogonal information from prompt $J/\psi$ pair hadroproduction instead.
In turn, this allows for a first experimental test of HQSS in heavy-quarkonium physics.

In fact, prompt $J/\psi$ pair hadroproduction provides a formidable laboratory to probe NRQCD factorization because the hadronization of $c\bar{c}$ pairs appears there twice thus providing enhanced sensitivity to the COM~\cite{Barger:1995vx,Qiao:2002rh,Li:2009ug,Ko:2010xy}.
Moreover, this final state is predicted to be frequently produced through double parton scattering (DPS) as well, which provides valuable access to the key parameter $\sigma_{\mathrm{eff}}$ of DPS~\cite{Kom:2011bd}.
Last but not least, prompt $J/\psi$ pair hadroproduction constitutes an irreducible background to dominant decays of charmed tetraquark states, such as the exotic $X(6900)$ hadron discovered some time ago as a resonance in the $m^{\psi\psi}$ spectrum \cite{LHCb:2020bwg}.
To date, prompt $J/\psi$ pair hadroproduction has been intensively studied by all four LHC collaborations: LHCb~\cite{LHCb:2020bwg,LHCb:2011kri,LHCb:2016wuo,LHCb:2023ybt}, CMS~\cite{CMS:2014cmt}, ATLAS~\cite{ATLAS:2016ydt}, and ALICE~\cite{ALICE:2023lsn}.
On the theoretical side, the first complete leading-order (LO) analysis in the collinear parton model (CPM), including all possible $\binom{8}{2}=28$ $(m,n)$ pairings, was performed in Ref~\cite{He:2015qya} and later improved in Ref.~\cite{He:2019qqr} by the inclusion of multiple initial-state gluon radiation via the Parton Reggeization Approach (PRA) \cite{Kniehl:2014qva,Karpishkov:2017kph} and the Balitsky-Fadin-Kuraev-Lipatov \cite{Kuraev:1976ge,Balitsky:1978ic} resummation of large logarithms in rapidity separation $|\Delta y^{\psi\psi}|$.
Apart from that, the study of higher-order corrections has so far been confined to the CS channel $gg\to2c\bar{c}({}^3\!S_1^{[1]})$, for which both quantum~\cite{Baranov:2012re,Lansberg:2013qka,Sun:2014gca,Lansberg:2014swa,Lansberg:2015lva,Lansberg:2019fgm,Sun:2023exa,Sun:2023exb,He:2025kkw} and relativistic~\cite{Li:2013csa,He:2024ugx} corrections have been considered.
For a review, we refer to Ref.~\cite{He:2021oyy}.

Comparing NLO CS predictions, including both $\mathcal{O}(\alpha_s)$~\cite{He:2025kkw} and $\mathcal{O}(v^2)$~\cite{He:2024ugx} corrections, with experiment, the LHCb measurements \cite{LHCb:2011kri,LHCb:2023ybt} were found to be well described in most bins, except for the kinematic end point regions, where the fixed-order treatment is invalidated by the radiation of soft gluons or hard collinear gluons and light quarks.
In the CMS \cite{CMS:2014cmt} and ATLAS \cite{ATLAS:2016ydt} cases, however, the NLO predictions turned out to undershoot the total cross sections by about one order of magnitude, despite sizable QCD correction factors of about 3, and to poorly describe the various cross section distributions \cite{He:2025kkw}.
The NLO predictions fall particularly short of the experimental data for large values of $p_T^{\psi\psi}$, $m^{\psi\psi}$, and $|\Delta y^{\psi\psi}|$ \cite{He:2025kkw}, and it is hopeless to expect these gaps to be filled by DPS contributions alone, even with a much smaller value of $\sigma_{\mathrm{eff}}$~\cite{ATLAS:2016ydt,Lansberg:2014swa,D0:2014vql}.
The quantitative difference between the LHCb case on the one hand and the ATLAS and CMS cases on the other hand may be traced to the different $p_T^{J/\psi}$ acceptance cuts imposed on each $J/\psi$ meson, namely, $p_T^{J/\psi}<10$~GeV \cite{LHCb:2011kri} and 14~GeV~\cite{LHCb:2023ybt} versus $p_T^{J/\psi}>4.5$~GeV~\cite{CMS:2014cmt} and 8.5~GeV~\cite{ATLAS:2016ydt}.

It is reasonable to expect that the above-mentioned gaps between NLO CS predictions and ATLAS and CMS data can be largely filled by taking the COM into account.
In the CPM framework, this would imply pushing the treatment of the 27 residual $(m,n)$ channels to NLO in $\alpha_s$ and $v^2$.
Besides the sheer magnitude of this task, conceptual problems related to NRQCD factorization breaking for double $P$-wave pairings are known to be lurking there \cite{He:2018hwb}.
As in the CS case \cite{He:2025kkw}, the $\mathcal{O}(\alpha_s)$ corrections arise from virtual and real radiation, with $2\to2$ and $2\to3$ kinematics, respectively.
To make a start, one could study real radiation in isolated mode, by imposing appropriate cuts on $J/\psi$ kinematic variables, e.g., a minimum-$p_T^{\psi\psi}$ cut.
In this way, one would also avoid dealing with NRQCD factorization breaking \cite{He:2018hwb}.

The importance of the COM at large values of $p_T^{\psi\psi}$, $m^{\psi\psi}$, and $|\Delta y^{\psi\psi}|$ has already been probed in the PRA framework \cite{He:2019qqr}, where $2\to3$ kinematics is mimicked by $k_T$ kicks in the initial state.
As already emphasized in Ref.~\cite{He:2019qqr}, meaningful extractions of $\sigma_{\mathrm{eff}}$ from experimental data of prompt $J/\psi$ pair hadroproduction require a complete NRQCD treatment including COM effects.
This is likely to be beneficial in resolving the notorious inconsistencies in $\sigma_{\mathrm{eff}}$ determinations from different sources \cite{ATLAS:2016ydt,LHCb:2023ybt,Lansberg:2014swa}. 

Looking at the CMS \cite{CMS:2014cmt} and ATLAS \cite{ATLAS:2016ydt} measurements, it is crucial to observe that the bulk of the cross section stems from the large-$p_T^{\psi\psi}$ region. 
In fact, a portion of approximately 35\% (70\%) of the CMS (ATLAS) cross section is sampled in the region of $p_T^{\psi\psi}>14$~GeV (15~GeV).
In these regions, there is a total of 17 experimental bins, which are sufficient in number for a detailed quantitative NRQCD analysis, which is precisely the purpose of this letter.

\begin{figure}
	\centering
	\begin{tabular}{c}
		\includegraphics[scale=0.30]{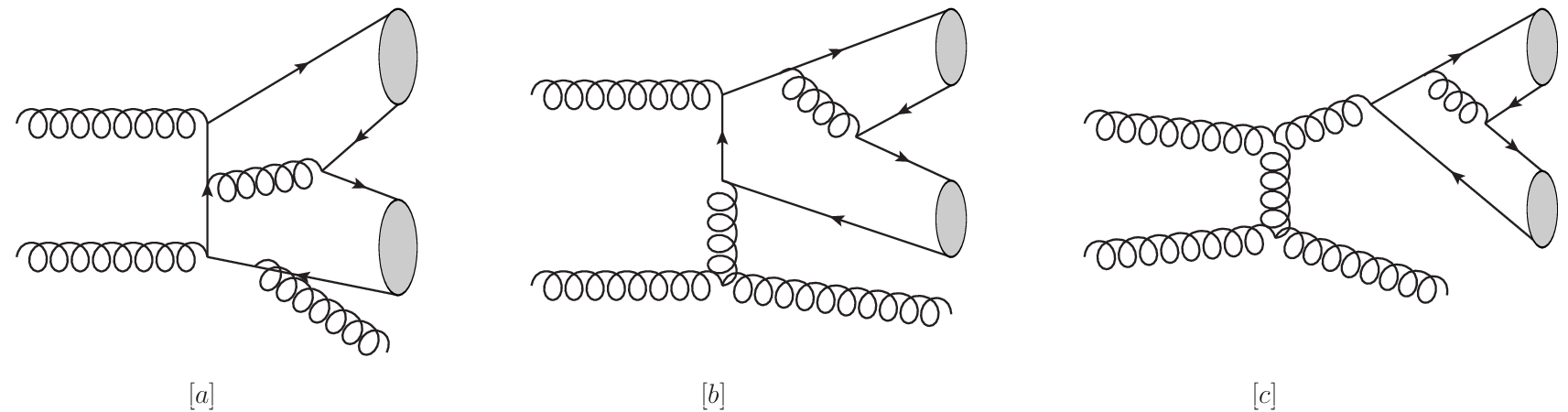}
	\end{tabular}
	\caption{Typical Feynman diagrams for $gg\to c\bar{c}(m)+c\bar{c}(n)+g$:
		(a) non-fragmentation; 
		(b) double parton fragmentation;
		(c) single parton fragmentation.}
	\label{Feynman}
\end{figure}

Invoking the CPM and NRQCD factorization, the hadronic cross section for prompt $J/\psi$ pair production in collisions of hadrons $A$ and $B$ reads:
\begin{eqnarray}\label{xs1}
 &&d\sigma\left(AB\to 2J/\psi+X\right) =\sum_{i,j,m,n,H_1,H_2}
  \int dx_1 d x_2
\nonumber\\
&&{}\times
f_{i/A}(x_1)f_{j/B}(x_2) d\hat{\sigma}(ij\to c\bar{c}(m)+c\bar{c}(n)+X)
\nonumber\\
&&{}\times
\langle\mathcal{O}^{H_1}(m)\rangle\langle\mathcal{O}^{H_2}(n)\rangle
\mathrm{Br}_{H_1\to J/\psi+X_1}\mathrm{Br}_{H_2\to J/\psi+X_2}\,,\ \ \ \
\end{eqnarray}
with parton density functions (PDFs) $f_{i/A}(x,\mu_f)$, SDCs $d\hat{\sigma}\left(ij\to c\bar{c}(m)+c\bar{c}(n)+X\right)$, LDMEs $\langle\mathcal{O}^H(m)\rangle$, and branching fractions $\mathrm{Br}(H\to J/\psi+X)$ of $H=J/\psi,\chi_{cJ},\psi^\prime$, being unity if $H=J/\psi$.
We work at LO in the fixed-flavor number scheme with incoming partons $i=g,q,\bar{q}$ ($q=u,d,s$) in all possible pairings, $gg,gq,g\bar{q},q\bar{q}$.

Having completed the CS case of $m=n={}^3\!S_1^{[1]}$ in Ref.~\cite{He:2025kkw}, we are left with 27 $(m,n)$ pairings.
Unfortunately, there is no publicly available program package that is able to evaluate their SDCs.
We accomplish this using QGRAF~\cite{Nogueira:1991ex} to generate the Feynman diagrams and FORM~\cite{Vermaseren:2000nd} to treat the Dirac and SU(3) color algebras. 
The double $P$-wave channels are most challenging because the scattering amplitudes need to be differentiated sequentially wrt to the relative momenta of the two Fock states resulting in lengthy analytical expressions of a few gigabyte.
To achieve the precision goal of 1\% in numerical integration, a quadruple-precision C++ program cooperating with the Cuba library~\cite{Hahn:2004fe} is developed.

\begin{table}
  \caption{Scaling with $p_T^{\psi\psi}$ of cross section $d\sigma/d(p_T^{\psi\psi})^2$ of $pp\to c\bar{c}(m)+c\bar{c}(n)+X$, dressed by respective LDMEs and branching fractions, for the various pairings $(m,n)$.}
	\begin{tabular}{|c|c|c|c|c|c|}
		\hline
		$(m,n)$ & ${}^3\!S_1^{[1]}$ & ${}^3\!S_1^{[8]}$ & ${}^1\!S_0^{[8]}$ &
		${}^3\!P_J^{[8]}$ & ${}^3\!P_J^{[1]}$ \\
		\hline
		${}^3\!S_1^{[1]}$ & $1/p_T^6$ & $v^4/p_T^6$ & $v^3/p_T^4$ & $v^4/p_T^4$ &  $v^4/p_T^8$ \\
		\hline
		${}^3\!S_1^{[8]}$ & $\cdots$ & $v^8/p_T^4$ & $v^7/p_T^4$ & $v^8/p_T^4$ & $v^8/p_T^4$ \\
		\hline
		${}^1\!S_0^{[8]}$ & $\cdots$ & $\cdots$ & $v^6/p_T^4$ & $v^7/p_T^4$ & $v^7/p_T^6$ \\
		\hline
		${}^3\!P_J^{[8]}$ & $\cdots$ & $\cdots$ & $\cdots$ & $v^8/p_T^4$ & $v^8/p_T^6$ \\
		\hline
		${}^3\!P_J^{[1]}$ & $\cdots$ & $\cdots$ & $\cdots$ & $\cdots$ & $v^8/p_T^6$ \\
		\hline
	\end{tabular}\label{p_T}
\end{table}

In the following discussion, we suppress $m={}^3\!P_0^{[1]}$ due to the smallness of $\mathrm{Br}(\chi_{c0}\to J/\psi\gamma)=(1.41\pm0.09)\%$ \cite{ParticleDataGroup:2024cfk}, so that we are left with $\binom{7}{2}=21$ channels.
In the large-$p_T^{\psi\psi}$ limit, their SDCs scale as $d\hat{\sigma}/d(p_T^{\psi\psi})^2\propto1/(p_T^{\psi\psi})^N+\mathcal{O}(1/(p_T^{\psi\psi})^{N+2})$, with $N$ depending on the channel considered because of $J^{PC}$ and color conservation.
According to the perturbative-QCD power counting rule developed in Ref.~\cite{Kang:2011mg}, we can divide the channels into three categories:
(i) Leading power (LP) with $N=4$, in which the two Fock states can be produced through both single and double parton fragmentation.
This includes ${}^3\!S_1^{[1]}+({}^1\!S_0^{[8]},{}^{3}\!P_J^{[8]})$, ${}^3\!P_J^{[1]}+{}^3\!S_1^{[8]}$, and all the double CO channels.
(ii) Next-to-leading power (NLP) with $N=6$, in which the two Fock states can only be produced through double parton fragmentation.
This includes the ${}^3\!S_1^{[1]}+({}^3\!S_1^{[1]},{}^3\!S_1^{[8]})$ and ${}^3\!P_J^{[1]}+({}^1\!S_0^{[8]},{}^3\!P_J^{[1,8]})$ channels.
(iii) Next-to-next-to-leading power (NNLP) with $N=8$, in which the two Fock states can neither be produced through single nor double parton fragmentation.
It only includes the CS $^3S_1^{[1]}+{}^3P_J^{[1]}$ channels.
Representative single parton, double parton, and non-fragmentation type Feynman 
diagrams of the $gg$ fusion sub-process are shown in Fig.~\ref{Feynman}. 
Together with the LDME velocity scaling rules \cite{Lepage:1992tx} and noticing that $\mathrm{Br}(\chi_{c1,2}\to J/\psi\gamma)$ are numerically of $\mathcal{O}(v^2)$, we can then roughly estimate the relative importance of each channel in the large-$p_T^{\psi\psi}$ limit; see Table~\ref{p_T}.

\begin{figure*}
	\centering
	\begin{tabular}{ccc}
		\includegraphics[scale=0.35]{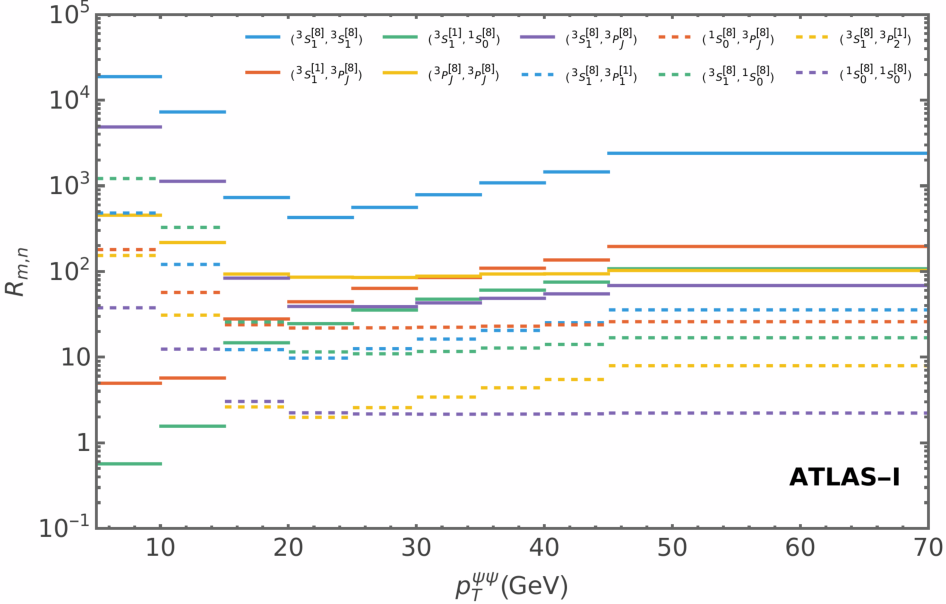}&
		\includegraphics[scale=0.35]{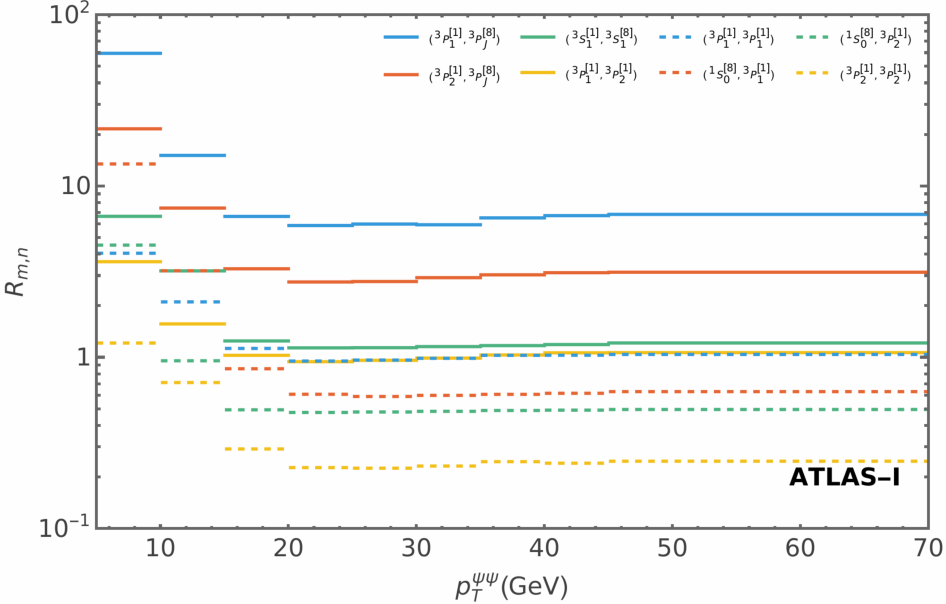}&
		\includegraphics[scale=0.35]{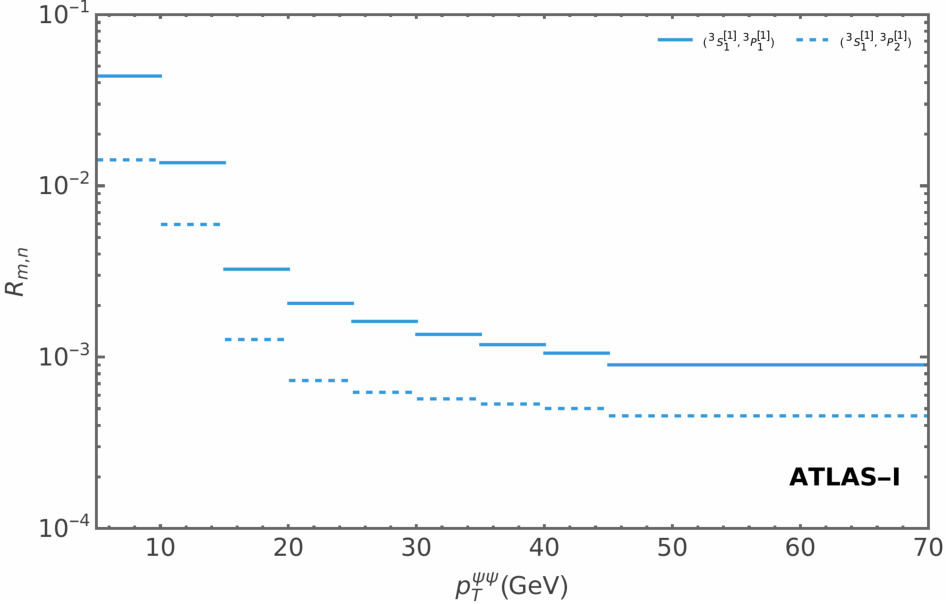}
	\end{tabular}
	\caption{Ratios $R_{m,n}$ in Eq.~\eqref{xs3} sorted by (a) LP, (b) NLP, and (c) NNLP for the ATLAS-I \cite{ATLAS:2016ydt} setup.}
	\label{ratios}
\end{figure*}

In our numerical analysis, we take the on-shell mass to be $m_c=1.5$~GeV, adopt the LO CTEQ6L1 PDF set \cite{Pumplin:2002vw}, and evaluate $\alpha^{(n_f)}_s(\mu_r)$ with $n_f=4$ and asymptotic scale parameter $\Lambda^{(4)}_{\mathrm{QCD}}=215$~MeV \cite{Pumplin:2002vw}.
We choose the renormalization and factorization scales to be $\mu_r=\mu_f=\xi\{[(p_T^{\psi_1})^2+4m_c^2]^{1/2}+[(p_T^{\psi_2})^2+4m_c^2]^{1/2}\}/2$, and vary $\xi$ from $1/2$ to 2 about its default value 1 to estimate theoretical uncertainties due to the lack of higher orders.
We use $\langle\mathcal{O}^{J/\psi}({}^3\!S_1^{[1]})\rangle=1.16~\mathrm{GeV}^3$ and $\langle\mathcal{O}^{\chi_{cJ}}({}^3\!P_J^{[1]})\rangle/(2J+1)=0.107~\mathrm{GeV}^5$ corresponding to the evaluation of the wave function at the origin in the Buchm\"uller-Tye potential~\cite{Eichten:1995ch}, and
$\langle\mathcal{O}^{\chi_{cJ}}({}^3\!S_1^{[8]})\rangle/(2J+1)=2.2\times10^{-3}~\mathrm{GeV}^5$ \cite{Ma:2010vd}.
In single prompt $J/\psi$ production, the feed-down contributions from $\chi_{cJ}$ and $\psi(2S)$ mesons are about $30\%$ and $10\%$, respectively \cite{Lansberg:2014swa}.
This implies that the $\psi(2S)$ feed-down contribution can be approximately accommodated by multiplying the hadronic cross section with a factor of $1+10\%/60\%=7/6$ for each direct $J/\psi$ meson in the final state, which is in line with
a recent LHCb measurement of prompt $J/\psi+\psi(2S)$ associated hadroproduction \cite{LHCb:2023wsl}.

CMS \cite{CMS:2014cmt} measured at $\sqrt{S}=7$~TeV and imposed $|y^{J/\psi}|<2.2$ and a $y^{J/\psi}$-dependent maximum-$p_T^{J/\psi}$ cut, ranging between 4.5~GeV and 6.5~GeV, on both $J/\psi$ mesons.
The DPS contamination of the CMS data was found \cite{Lansberg:2014swa} to range at the few-percent level in the $p_T^{\psi\psi}$ range selected for our analysis and can safely be neglected.
ATLAS \cite{ATLAS:2016ydt} measured at $\sqrt{S}=8$~TeV, imposed $p_T^{J/\psi}>8.5$~GeV and $|y^{J/\psi}|<2.1$ on both $J/\psi$ mesons, and discriminated between (I) central ($|y^{J/\psi_2}|<1.05$) and (II) forward ($1.05<|y^{J/\psi_2}|<2.1$) subleading $J/\psi$ meson, with $p_T^{J/\psi_2}<p_T^{J/\psi_1}$.
ATLAS managed to extract the DPS contribution, to be subtracted for our purposes.

To quantitatively verify the $p_T^{\psi\psi}$ scaling rules of Table~\ref{p_T}, we evaluate the SDCs for the 20 non-CS $(m,n)$ pairs, normalized and rendered dimensionless as
\begin{eqnarray}\label{xs3}
R_{m,n}=\frac{d\hat{\sigma}(pp\to 
c\bar{c}(m)+c\bar{c}(n)+X)\,m_c^{d_m+d_n-6}}{d\hat{\sigma}(pp\to 
2c\bar{c}({}^3\!S_1^{[1]})+X)}\,,
\end{eqnarray}
where $d_m$ is the mass dimension of $\mathcal{O}^H(m)$, for the exemplary case of the ATLAS-I setup and show the results separately for the LP, NLP, and NNLP channels in Fig.~\ref{ratios}(a)--(c), respectively.
The results for the ATLAS-II and CMS setups are very similar.
As expected, we observe from Fig.~\ref{ratios} that, for sufficiently large $p_T^{\psi\psi}$, $R_{m,n}$ scales as $(p_T^{\psi\psi})^N$ with $N=2,0,-2$ for LP, NLP, and NNLP channels $(m,n)$, with the exception of the LP channels $({}^1\!S_0^{[8]},{}^1\!S_0^{[8]})$, $({}^1\!S_0^{[8]},{}^3\!P_J^{[8]})$, and $({}^3\!P_J^{[8]},{}^3\!P_J^{[8]})$.
This exceptional behavior may be understood by observing that the coupling between the fragmenting gluon and the Fock states $m$ and $n$ is proportional to the relative momentum between the latter, which yields the overall factor $(M_{m,n}^2-16m_c^2)$ in the squared scattering amplitude, where $M_{m,n}$ is the invariant mass of the $(m,n)$ pair.
In the fragmentation limit, $M_{m,n}$ approaches $16m_c^2$ so that the contribution due to single parton fragmentation is effectively quenched, making double parton fragmentation prominent.
There is a clear hierarchy in size between the LP, NLP, and NNLP classes in Figs.~\ref{ratios}(a)--(c).
$R_{{}^3\!S_1^{[8]},{}^3\!S_1^{[8]}}$ is largest within the LP class.
However, this is suppressed in the hadronic cross section of Eq.~\eqref{xs1} by two powers of $\mathcal{O}^{J/\psi}({}^3\!S_1^{[8]})$, whose magnitude was generally found to be below $2\times10^{-2}~\mathrm{GeV}^3$ \cite{Ma:2010yw,Butenschoen:2010rq,Butenschoen:2011yh,Chao:2012iv,Gong:2012ug,Bodwin:2014gia,Shao:2014yta,Brambilla:2024iqg}.
Similar comments apply to the other pure CO LP channels.
On the other hand, the mixed LP channels ${}^3\!S_1^{[8]}+{}^3\!P_J^{[1]}$ are suppressed by the smallness of $\langle\mathcal{O}^{\chi_{cJ}}({}^3\!P_J^{[1]})\rangle$ and $\mathrm{Br}(\chi_{cJ}\to J/\psi\gamma)$.
As predicted by the superficial power counting rules in Table~\ref{p_T}, the ${}^3\!S_1^{[1]}+({}^1\!S_0^{[8]},{}^3\!P_J^{[8]})$ channels indeed greatly dominate the LP class in Fig.~\ref{ratios}(a).
Furthermore, we observe that the $R_{{}^3\!S_1^{[1]},{}^3\!P_J^{[8]}}$ to $R_{{}^3\!S_1^{[1]},{}^1\!S_0^{[8]}}$ ratio is approximately constant in the large-$p_T^{\psi\psi}$ region.
Considering the CMS bins with $p_T^{\psi\psi}>18$~GeV and the ATLAS-I and ATLAS-II bins with $p_T^{\psi\psi}>20$~GeV, which are 14 ones altogether, this ratio ranges from 1.77 to 1.86, the average value being 1.80.
This implies that $\langle\mathcal{O}^{J/\psi}({}^1\!S_0^{[8]})\rangle$ and $\langle\mathcal{O}^{J/\psi}({}^3\!P_0^{[8]})\rangle$ are not separately constrained by these bins, but only their linear combination $M_{0,1.8}^{J/\psi}$ is, where \cite{Ma:2010yw,Shao:2014yta}
\begin{eqnarray}
  M_{0,r_0}^{J/\psi}&=&\langle\mathcal{O}^{J/\psi}({}^1\!S_0^{[8]})\rangle+\frac{r_0}{m_c^2}\langle\mathcal{O}^{J/\psi}(^3P_0^{[8]})\rangle\,,
  \nonumber\\
  M_{1,r_1}^{J/\psi}&=&\langle\mathcal{O}^{J/\psi}({}^3\!S_1^{[8]})\rangle+\frac{r_1}{m_c^2}\langle\mathcal{O}^{J/\psi}(^3P_0^{[8]})\rangle\,.
\end{eqnarray}
Within the NLP class, the pure CS channel $2\,{}^3\!S_1^{[1]}$ is by far dominant and competes in size with the ${}^3\!S_1^{[1]}+({}^1\!S_0^{[8]},{}^3\!P_J^{[8]})$ channels.

These observations motivate the following strategy for fitting the high-$p_T^{\psi\psi}$ CMS \cite{CMS:2014cmt}, ATLAS-I and ATLAS-II \cite{ATLAS:2016ydt} data:
(i) include only the ${}^3\!S_1^{[1]}+({}^3\!S_1^{[1]},{}^1\!S_0^{[8]})$ channels and fit $M_{0,1.8}^{J/\psi}$ to the 14 bins with $p_T^{\psi\psi}>18$~GeV;
(ii) combine this with $M_{0,3.9}^{J/\psi}$ and $M_{1,-0.56}^{J/\psi}$ from a fit to single prompt $J/\psi$ hadroproduction \cite{Ma:2010yw,Shao:2014yta}, including LP and NLP channels \cite{Kang:2014tta,Ma:2014svb}, and so pin down
$\langle\mathcal{O}^{J/\psi}({}^1\!S_0^{[8]})\rangle$,
$\langle\mathcal{O}^{J/\psi}({}^3\!S_1^{[8]})\rangle$, and 
$\langle\mathcal{O}^{J/\psi}({}^3\!P_0^{[8]})\rangle$ separately;
(iii) verify that $\chi^2/\mathrm{dof}$ is stable when all 21 channels are included.
The fit in step~(i) yields $\chi^2/\mathrm{dof}=1.31$, which is mostly built up by the last two ATLAS-I bins, with $p_T^{\psi\psi}>40$~GeV, where the data systematically undershoot the theoretical evaluations.
Excluding them dramatically improves the fit, yielding $\chi^2/\mathrm{dof}=0.38$ along with $M_{0,1.8}^{J/\psi}=(3.49\pm0.59)\times10^{-2}~\mathrm{GeV}^3$. 
Combining this with $M_{0,3.9}^{J/\psi}=(7.4\pm1.9)\times10^{-2}~\mathrm{GeV}^{3}$ and $M_{1,-0.56}^{J/\psi}=(0.05\pm0.02)\times10^{-2}~\mathrm{GeV}^{3}$ \cite{Ma:2010yw,Shao:2014yta}, we obtain the new $J/\psi$ CO LDME set in Table~\ref{fit}.
Re-evaluating $\chi^2/\mathrm{dof}$ with all in, we obtain 0.40, being just 5\% larger than the fit score, which ultimately consolidates our fit philosophy in hindsight.

\begin{table}
  \caption{Our fit results (up) and those of Ref.~\cite{Brambilla:2024iqg} for central scale choice (down) in units of $10^{-2}~\mathrm{GeV}^{3}$.}
	\begin{tabular}{ccc}
	  \hline
          $\langle\mathcal{O}^{J/\psi}({}^1\!S_0^{[8]})\rangle$ &
          $\langle\mathcal{O}^{J/\psi}({}^3\!S_1^{[8]})\rangle$ &
          $\langle\mathcal{O}^{J/\psi}({}^3\!P_0^{[8]})\rangle/m_c^2$ \\
          \hline
          $0.14\pm1.96$ & $1.09\pm0.53$ & $1.86\pm0.95$ \\
          $0.068\pm0.249$ & $1.05\pm0.12$ & $1.88\pm0.26$ \\  
          \hline
	\end{tabular}\label{fit}
\end{table}

\begin{figure*}
	\centering
	\begin{tabular}{ccc}
		\includegraphics[width=0.33\linewidth]{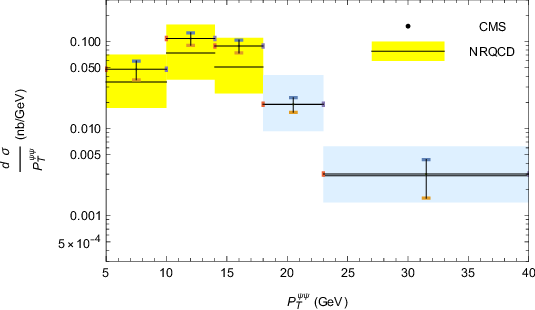}&
		\includegraphics[width=0.33\linewidth]{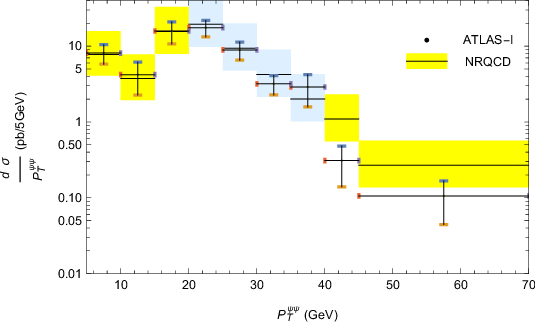}&
		\includegraphics[width=0.33\linewidth]{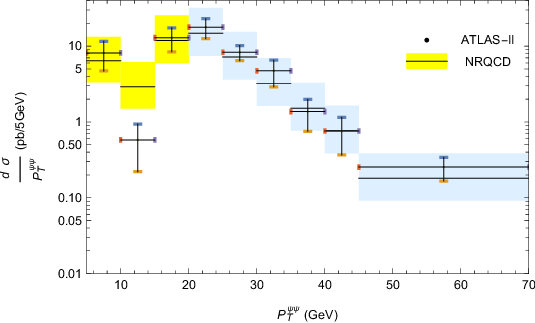}\\
		\includegraphics[width=0.33\linewidth]{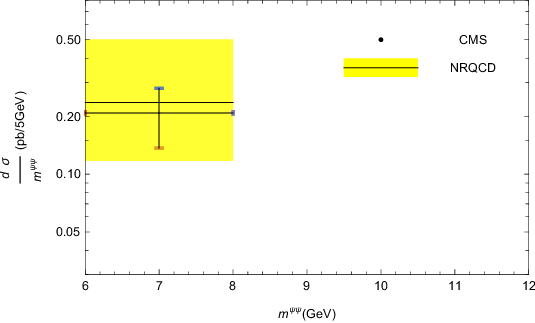}&
		\includegraphics[width=0.33\linewidth]{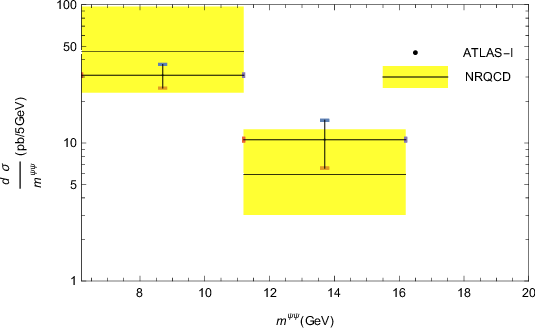}&
		\includegraphics[width=0.33\linewidth]{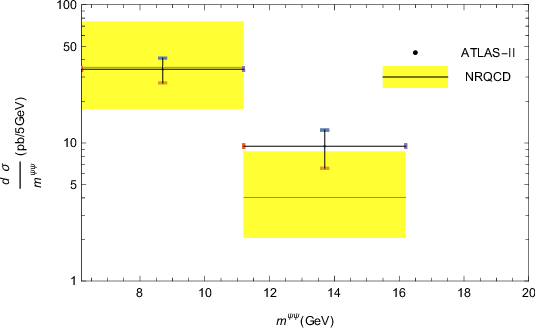}\\
	\end{tabular}
	\caption{CMS \cite{CMS:2014cmt} (left), ATLAS-I (center), and ATLAS-II \cite{ATLAS:2016ydt} (right) measurements of $d\sigma/p_T^{\psi\psi}$ (up) and
          $d\sigma/m^{\psi\psi}$ (down) are compared with full $\mathcal{O}(\alpha_s^5)$ NRQCD.
          Theoretical uncertainties are indicated by shaded (colored) bands.
Dark (blue) bins are included in the fit; light (yellow) bins correspond to genuine predictions.}
	\label{ptjj}
\end{figure*}

The goodness of our fit is nicely illustrated in Figs.~\ref{ptjj}(a)--(c), where the bins included in the fit are marked by dark (blue) bands.
It is reassuring that also the excluded data points are very well described, except for the eighth ATLAS-I and second ATLAS-II bins, where experimental and theoretical error bars do not overlap, the gaps being tiny though.
Our $\mathcal{O}(\alpha_s^5)$ predictions also nicely agree with the data in the lowest CMS, ATLAS-I, and ATLAS-II $m^{\psi\psi}$ bins, outside $\mathcal{O}(\alpha_s^4)$ kinematics, as is evident from Figs.~\ref{ptjj}(d)--(f).

A preliminary assessment of how well $\mathcal{O}(\alpha_s^5)$ evaluations with LDME sets from other groups agree with CMS \cite{CMS:2014cmt} and ATLAS \cite{ATLAS:2016ydt} data at large $p_T^{\psi\psi}$ may be gained by considering $M_{0,1.8}^{J/\psi}$.
The most frequently used alternative sets \cite{Butenschoen:2011yh,Gong:2012ug,Bodwin:2014gia} have $M_{0,1.8}^{J/\psi}=\{2.3\pm0.37,8.0\pm1.0,10.8\pm2.3\}\times10^{-2}~\mathrm{GeV^{3}}$ and thus yield central predictions that correspondingly undershoot or overshoot the data in the fit ranges, with $\chi^2/\mathrm{dof}=\{0.44,5.6,14.8\}$, respectively.
Detailed inspection reveals that experimental and theoretical errors always overlap in the fit ranges.
However, the predictions based on Refs.~\cite{Butenschoen:2011yh,Gong:2012ug} fall short of the ATLAS-I and ATLAS-II data in the first bin, which may be traced to the relatively small values of $\langle\mathcal{O}^{J/\psi}({}^3\!S_1^{[8]})\rangle$.

Incidentally, our $J/\psi$ LDME set is very close to the one obtained in Ref.~\cite{Brambilla:2024iqg} for central scale choice $\mu_r=\mu_f=[(p_T^{J/\psi})^2+4m_c^2]^{1/2}$, which is also listed in Table~\ref{fit}.
There is some difference in the central values of $\langle\mathcal{O}^{J/\psi}({}^1\!S_0^{[8]})\rangle$, which is largely inconsequential---except for the $\eta_c$ case discussed below---, however, because they are both compatible with zero.
This implies that the good overall agreement with the world data of prompt $J/\psi$ yield and polarization observed in Ref.~\cite{Brambilla:2024iqg} carries over to our analysis. 
The fit strategy of Ref.~\cite{Brambilla:2024iqg} is outlined above.
It crucially differs from ours in that it is built upon the HQSS hypothesis, while we fully rely on experimental data of prompt $J/\psi$ hadroproduction, both singly and in pairs.
In other words, our $\mathcal{O}(\alpha_s^5)$ analysis of the CMS \cite{CMS:2014cmt} and ATLAS \cite{ATLAS:2016ydt} data at large $p_T^{\psi\psi}$ provides a first genuine verification of HQSS in heavy quarkonium.
Comparing NLO predictions for prompt $\eta_c$ yield \cite{Butenschoen:2014dra,Brambilla:2024iqg} with LHCb data at $\sqrt{s}=7,8,13$~TeV \cite{LHCb:2014oii,LHCb:2019zaj}, we obtain $\chi^2/11=\{1.52,1.38\}$ for the two LDME sets in Table~\ref{fit}, the small difference being due to $\langle\mathcal{O}^{J/\psi}({}^1\!S_0^{[8]})\rangle$.

In summary, we performed the first complete $\mathcal{O}(\alpha_s^5)$ NRQCD analysis of prompt $J/\psi$ pair hadroproduction in the large-$p_T^{\psi\psi}$ region, classified the 28 contributing channels in LP, NLP, and NNLP, and explored their relative magnitudes.
We thus revealed that CMS \cite{CMS:2014cmt} and ATLAS \cite{ATLAS:2016ydt} data constrain the linear combination $M_{0,1.8}^{J/\psi}$ of LDMEs.
Combining $M_{0,1.8}^{J/\psi}$ with $M_{0,3.9}^{J/\psi}$ and $M_{1,-0.56}^{J/\psi}$ previously extracted from single prompt $J/\psi$ hadroproduction \cite{Ma:2010yw,Shao:2014yta}, we obtained a new LDME set, which happens to be very similar to that in Ref.~\cite{Brambilla:2024iqg} and describes well the world data of prompt $J/\psi$ yield and polarization and also the LHCb \cite{LHCb:2014oii,LHCb:2019zaj} data of prompt $\eta_c$ yield if HQSS is valid.
Unlike Ref.~\cite{Brambilla:2024iqg}, our fit only relies on $J/\psi$ data and thus provides a crucial test of HQSS.

We thank Mathias Butensch\"on and Yan-Qing Ma for beneficial discussions.
This work was supported in part by the German Research Foundation DFG through Grants No.~KN 365/13-2 and No.~KN 365/14-2, and by Fundamental Research Funds for
the Central Universities through Grant No.~buctrc202432.


\begin{thebibliography}{99}

\bibitem{Bodwin:1994jh}
G.~T.~Bodwin, E.~Braaten, and G.~P.~Lepage,
Rigorous QCD analysis of inclusive annihilation and production of heavy quarkonium,
Phys.\ Rev.\ D \textbf{51}, 1125--1171 (1995);
\textbf{55}, 5853(E) (1997)
[arXiv:hep-ph/9407339 [hep-ph]].

\bibitem{Caswell:1985ui}
W.~E.~Caswell and G.~P.~Lepage,
Effective lagrangians for bound state problems in QED, QCD, and other field
theories,
Phys.\ Lett.\ B \textbf{167}, 437--442 (1986).

\bibitem{Lepage:1992tx}
G.~P.~Lepage, L.~Magnea, C.~Nakhleh, U.~Magnea, and K.~Hornbostel,
Improved nonrelativistic QCD for heavy-quark physics,
Phys.\ Rev.\ D \textbf{46}, 4052--4067 (1992)
[arXiv:hep-lat/9205007 [hep-lat]].

\bibitem{Butenschoen:2022wld}
M.~Butenschoen and B.~A.~Kniehl,
Constraints on Nonrelativistic-QCD Long-Distance Matrix Elements from $J/\psi$
Plus $W/Z$ Production at the LHC,
Phys.\ Rev.\ Lett.\ \textbf{130}, 041901 (2023)
[arXiv:2207.09366 [hep-ph]].

\bibitem{QuarkoniumWorkingGroup:2004kpm}
N.~Brambilla \textit{et al.}\ (Quarkonium Working Group),
Heavy Quarkonium Physics,
CERN Yellow Report No.~CERN-2005-005
[arXiv:hep-ph/0412158 [hep-ph]].

\bibitem{Lansberg:2006dh}
J.-P.~Lansberg,
$J/\psi$, $\psi^\prime$ and $\Upsilon$ Production at Hadron Colliders: A Review,
Int.\ J. Mod.\ Phys.\ A \textbf{21}, 3857--3916 (2006)
[arXiv:hep-ph/0602091 [hep-ph]].

\bibitem{Brambilla:2010cs}
  N.~Brambilla
  \textit{et al.},
Heavy Quarkonium: progress, puzzles, and opportunities,
Eur.\ Phys.\ J. C \textbf{71}, 1534 (2011)
[arXiv:1010.5827 [hep-ph]].

\bibitem{Butenschoen:2012qr}
M.~Butenschoen and B.~A.~Kniehl,
Next-to-Leading-Order Tests of Non-Relativistic-QCD Factorization with $J/\psi$ Yield and Polarization,
Mod.\ Phys.\ Lett.\ A \textbf{28}, 1350027 (2013)
[arXiv:1212.2037 [hep-ph]].

\bibitem{Brambilla:2014jmp}
  N.~Brambilla
  \textit{et al.},
  QCD and strongly coupled gauge theories: challenges and perspectives,
Eur.\ Phys.\ J. C \textbf{74}, 2981 (2014)
[arXiv:1404.3723 [hep-ph]].

\bibitem{Lansberg:2019adr}
J.-P.~Lansberg,
New observables in inclusive production of quarkonia,
Phys.\ Rept.\ \textbf{889}, 1--106 (2020)
[arXiv:1903.09185 [hep-ph]].

\bibitem{Ma:2010yw}
Y.-Q.~Ma, K.~Wang, and K.-T.~Chao,
$J/\psi$ ($\psi^\prime$) Production at the Tevatron and LHC at ${\cal O}(\alpha_s^4v^4)$ in Nonrelativistic QCD,
Phys.\ Rev.\ Lett.\ \textbf{106}, 042002 (2011)
[arXiv:1009.3655 [hep-ph]].

\bibitem{Butenschoen:2010rq}
M.~Butensch\"on and B.~A.~Kniehl,
Reconciling $J/\psi$ Production at HERA, RHIC, Tevatron, and LHC with Nonrelativistic QCD Factorization at Next-to-Leading Order,
Phys.\ Rev.\ Lett.\ \textbf{106}, 022003 (2011)
[arXiv:1009.5662 [hep-ph]].

\bibitem{Butenschoen:2011yh}
M.~Butenschoen and B.~A.~Kniehl,
World data of $J/\psi$ production consolidate nonrelatvistic QCD factorization at next-to-leading order,
Phys.\ Rev.\ D \textbf{84}, 051501(R) (2011)
[arXiv:1105.0820 [hep-ph]].

\bibitem{Chao:2012iv}
K.-T.~Chao, Y.-Q.~Ma, H.-S.~Shao, K.~Wang, and Y.-J.~Zhang,
$J/\psi$ Polarization at Hadron Colliders in Nonrelativistic QCD,
Phys.\ Rev.\ Lett.\ \textbf{108}, 242004 (2012)
[arXiv:1201.2675 [hep-ph]].

\bibitem{Gong:2012ug}
B.~Gong, L.-P.~Wan, J.-X.~Wang, and H.-F.~Zhang,
Polarization for Prompt $J/\psi$ and $\psi(2s)$ Production at the Tevatron and LHC,
Phys.\ Rev.\ Lett.\ \textbf{110}, 042002 (2013)
[arXiv:1205.6682 [hep-ph]].

\bibitem{Bodwin:2014gia}
G.~T.~Bodwin, H.~S.~Chung, U-R.~Kim, and J.~Lee,
Fragmentation Contributions to $J/\psi$ Production at the Tevatron and the LHC,
Phys.\ Rev.\ Lett.\ \textbf{113}, 022001 (2014)
[arXiv:1403.3612 [hep-ph]].

\bibitem{Shao:2014yta}
H.-S.~Shao, H.~Han, Y.-Q.~Ma, C.~Meng, Y.-J.~Zhang, and K.-T.~Chao,
Yields and polarizations of prompt $J/\psi$ and $\psi(2S)$ production in hadronic collisions,
JHEP \textbf{05}, 103 (2015)
[arXiv:1411.3300 [hep-ph]].

\bibitem{Butenschoen:2009zy}
M.~Butensch\"on and B.~A.~Kniehl,
Complete Next-to-Leading-Order Corrections to $J/\psi$ Photoproduction in Nonrelativistic Quantum Chromodynamics,
Phys.\ Rev.\ Lett.\ \textbf{104}, 072001 (2010)
[arXiv:0909.2798 [hep-ph]].

\bibitem{Butenschoen:2011ks}
M.~Butenschoen and B.~A.~Kniehl,
Probing Nonrelativistic QCD Factorization in Polarized $J/\psi$ Photoproduction at Next-to-Leading Order,
Phys.\ Rev.\ Lett.\ \textbf{107}, 232001 (2011)
[arXiv:1109.1476 [hep-ph]].

\bibitem{Butenschoen:2012px}
M.~Butenschoen and B.~A.~Kniehl,
$J/\psi$ Polarization at the Tevatron and the LHC: Nonrelativistic-QCD Factorization at the Crossroads,
Phys.\ Rev.\ Lett.\ \textbf{108}, 172002 (2012)
[arXiv:1201.1872 [hep-ph]].

\bibitem{Butenschoen:2014dra}
M.~Butenschoen, Z.-G.~He, and B.~A.~Kniehl,
$\eta_c$ Production at the LHC Challenges Nonrelativistic QCD Factorization,
Phys.\ Rev.\ Lett.\ \textbf{114}, 092004 (2015)
[arXiv:1411.5287 [hep-ph]].

\bibitem{Butenschoen:2019lef}
M.~Butenschoen and B.~A.~Kniehl,
Dipole subtraction at next-to-leading order in nonrelativistic-QCD factorization,
Nucl.\ Phys.\ B \textbf{950}, 114843 (2020)
[arXiv:1909.03698 [hep-ph]].

\bibitem{Butenschoen:2020mzi}
M.~Butenschoen and B.~A.~Kniehl,
Dipole subtraction vs.\ phase space slicing in NLO NRQCD heavy-quarkonium production calculations,
Nucl.\ Phys.\ B \textbf{957}, 115056 (2020)
[arXiv:2003.01014 [hep-ph]].

\bibitem{Brambilla:2024iqg}
N.~Brambilla, M.~Butenschoen, and X.-P.~Wang,
How well does nonrelativistic QCD factorization work at next-to-leading order?,
Phys.\ Rev.\ D \textbf{112}, L011902 (2025)
[arXiv:2411.16384 [hep-ph]].

\bibitem{CMS:2015lbl}
V.~Khachatryan \textit{et al.}\ (CMS Collaboration),
Measurement of $J/\psi$ and $\psi(2S)$ Prompt Double-Differential Cross Sections in $pp$ Collisions at $\sqrt{s}=7$~TeV,
Phys.\ Rev.\ Lett.\ \textbf{114} (2015), 191802
[arXiv:1502.04155 [hep-ex]].

\cite{LHCb:2014oii}
\bibitem{LHCb:2014oii}
R.~Aaij \textit{et al.}\ (LHCb Collaboration),
Measurement of the $\eta_c(1S)$ production cross-section in proton--proton collisions via the decay $\eta_c(1S) \rightarrow p\bar{p}$,
Eur.\ Phys.\ J.\ C \textbf{75}, 311 (2015)
[arXiv:1409.3612 [hep-ex]].

\bibitem{LHCb:2019zaj}
R.~Aaij \textit{et al.}\ (LHCb Collaboration),
Measurement of the $\eta_c(1S)$ production cross-section in $pp$ collisions at $\sqrt{s} = 13$~TeV,
Eur.\ Phys.\ J. C \textbf{80}, 191 (2020)
[arXiv:1911.03326 [hep-ex]].


\bibitem{Han:2014jya}
H.~Han, Y.-Q.~Ma, C.~Meng, H.-S.~Shao, and K.-T.~Chao,
$\eta_c$ Production at LHC and Indications on the Understanding of $J/\psi$ Production,
Phys.\ Rev.\ Lett.\ \textbf{114}, 092005 (2015)
[arXiv:1411.7350 [hep-ph]].

\bibitem{Zhang:2014ybe}
H.-F.~Zhang, Z.~Sun, W.-L.~Sang, and R.~Li,
Impact of $\eta_c$ Hadroproduction Data on Charmonium Production and Polarization within the Nonrelativistic QCD Framework,
Phys.\ Rev.\ Lett.\ \textbf{114}, 092006 (2015)
[arXiv:1412.0508 [hep-ph]].

\bibitem{Barger:1995vx}
V.~Barger, S.~Fleming, and R.~J.~N.~Phillips,
Double gluon fragmentation to $J/\psi$ pairs at the Tevatron,
Phys.\ Lett.\ B \textbf{371}, 111--116 (1996)
[arXiv:hep-ph/9510457 [hep-ph]].

\bibitem{Qiao:2002rh}
C.-F.~Qiao,
$J/\psi$ pair production at the Fermilab Tevatron,
Phys.\ Rev.\ D \textbf{66}, 057504 (2002)
[arXiv:hep-ph/0206093 [hep-ph]].

\bibitem{Li:2009ug}
R.~Li, Y.-J.~Zhang. and K.-T.~Chao,
Pair production of heavy quarkonium and $B_c^{(*)}$ mesons at hadron colliders,
Phys.\ Rev.\ D \textbf{80}, 014020 (2009)
[arXiv:0903.2250 [hep-ph]].

\bibitem{Ko:2010xy}
P.~Ko, J.~Lee, and C.~Yu,
Inclusive double-quarkonium production at the Large Hadron Collider,
JHEP \textbf{01}, 070 (2011)
[arXiv:1007.3095 [hep-ph]].

\bibitem{Kom:2011bd}
C.~H.~Kom, A.~Kulesza, and W.~J.~Stirling,
Pair Production of $J/\psi$ as a Probe of Double Parton Scattering at LHCb,
Phys.\ Rev.\ Lett.\ \textbf{107}, 082002 (2011)
[arXiv:1105.4186 [hep-ph]].

\bibitem{LHCb:2020bwg}
R.~Aaij \textit{et al.}\ (LHCb Collaboration),
Observation of structure in the $J /\psi$-pair mass spectrum,
Sci.\ Bull.\ \textbf{65}, 1983--1993 (2020)
[arXiv:2006.16957 [hep-ex]].

\bibitem{LHCb:2011kri}
R.~Aaij \textit{et al.}\ (LHCb Collaboration),
Observation of $J/\psi$-pair production in $pp$ collisions at $\sqrt{s}=7$~TeV,
Phys.\ Lett.\ B \textbf{707}, 52--59 (2012)
[arXiv:1109.0963 [hep-ex]].

\bibitem{LHCb:2016wuo}
R.~Aaij \textit{et al.}\ (LHCb Collaboration),
Measurement of the $J/\psi$ pair production cross-section in $pp$ collisions at $\sqrt{s}=13$~TeV,
JHEP \textbf{06}, 047 (2017);
\textbf{10}, 068(E) (2017)
[arXiv:1612.07451 [hep-ex]].

\bibitem{LHCb:2023ybt}
R.~Aaij \textit{et al.}\ (LHCb Collaboration),
Measurement of $J/\psi$-pair production in $pp$ collisions at $\sqrt{s}=13$~TeV and study of gluon transverse-momentum dependent PDFs,
JHEP \textbf{03}, 088 (2024)
[arXiv:2311.14085 [hep-ex]].

\bibitem{CMS:2014cmt}
V.~Khachatryan \textit{et al.}\ (CMS Collaboration),
Measurement of prompt $J/\psi$ pair production in $pp$ collisions at $\sqrt{s}=7$~Tev,
JHEP \textbf{09}, 094 (2014)
[arXiv:1406.0484 [hep-ex]].

\bibitem{ATLAS:2016ydt}
M.~Aaboud \textit{et al.}\ (ATLAS Collaboration),
Measurement of the prompt $J/\psi$ pair production cross-section in $pp$ collisions at $\sqrt{s}=8$~TeV with the ATLAS detector,
Eur.\ Phys.\ J. C \textbf{77}, 76 (2017)
[arXiv:1612.02950 [hep-ex]].

\bibitem{ALICE:2023lsn}
S.~Acharya \textit{et al.}\ (ALICE Collaboration),
Measurement of inclusive $J/\psi$ pair production cross section in $pp$ collisions at $\sqrt{s}=13$~TeV,
Phys.\ Rev.\ C \textbf{108}, 045203 (2023)
[arXiv:2303.13431 [hep-ex]].

\bibitem{He:2015qya}
Z.-G.~He and B.~A.~Kniehl,
Complete Nonrelativistic-QCD Prediction for Prompt Double $J/\psi$ Hadroproduction,
Phys.\ Rev.\ Lett.\ \textbf{115}, 022002 (2015)
[arXiv:1609.02786 [hep-ph]].

\bibitem{He:2019qqr}
Z.-G.~He, B.~A.~Kniehl, M.~A.~Nefedov, and V.~A.~Saleev,
Double Prompt $J/\psi$ Hadroproduction in the Parton Reggeization Approach with High-Energy Resummation,
Phys.\ Rev.\ Lett.\ \textbf{123}, 162002 (2019)
[arXiv:1906.08979 [hep-ph]].

\bibitem{Kniehl:2014qva}
B.~A.~Kniehl, M.~A.~Nefedov, and V.~A.~Saleev,
Prompt-photon plus jet associated photoproduction at HERA in the parton Reggeization approach,
Phys.\ Rev.\ D \textbf{89}, 114016 (2014)
[arXiv:1404.3513 [hep-ph]].

\bibitem{Karpishkov:2017kph}
A.~V.~Karpishkov, M.~A.~Nefedov, and V.~A.~Saleev,
$B{\bar B}$ angular correlations at the LHC in parton Reggeization approach merged with higher-order matrix elements,
Phys.\ Rev.\ D \textbf{96}, 096019 (2017)
[arXiv:1707.04068 [hep-ph]].

\bibitem{Kuraev:1976ge}
E.~A.~Kuraev, L.~N.~Lipatov, and V.~S.~Fadin,
Multiregge processes in the Yang--Mills theory,
Zh.\ Eksp.\ Teor.\ Fiz.\ {\bf 71}, 840--855 (1976)
[Sov.\ Phys.\ JETP \textbf{44}, 443--451 (1976)].
  
\bibitem{Balitsky:1978ic} 
Ya.~Ya.~Balitski\u{i} and L.~N.~Lipatov,
The Pomeranchuk singularity in quantum chromodynamics,
Yad.\ Fiz.\ \textbf{28}, 1597--1611 (1978) 
[Sov.\ J. Nucl.\ Phys.\ {\bf 28}, 822--829 (1978)].

\bibitem{Baranov:2012re}
S.~P.~Baranov, A.~M.~Snigirev, N.~P.~Zotov, A.~Szczurek, and W.~Sch{\"a}fer,
Interparticle correlations in the production of $J/\psi$ pairs in proton-proton collisions,
Phys.\ Rev.\ D \textbf{87}, 034035 (2013)
[arXiv:1210.1806 [hep-ph]].

\bibitem{Lansberg:2013qka}
J.-P.~Lansberg and H.~S.~Shao,
Production of $J/\psi + \eta_{c}$ versus $J/\psi + J/\psi$ at the LHC: Importance of Real $\alpha^{5}_{s}$ Corrections,
Phys.\ Rev.\ Lett.\ \textbf{111}, 122001 (2013)
[arXiv:1308.0474 [hep-ph]].

\bibitem{Sun:2014gca}
L.-P.~Sun, H.~Han, and K.-T.~Chao,
Impact of $J/\psi$ pair production at the LHC and predictions in nonrelativistic QCD,
Phys.\ Rev.\ D \textbf{94}, 074033 (2016)
[arXiv:1404.4042 [hep-ph]].

\bibitem{Lansberg:2014swa}
J.-P.~Lansberg and H.-S.~Shao,
$J/\psi$-pair production at large momenta: Indications for double parton scatterings and large $\alpha_s^5$ contributions,
Phys.\ Lett.\ B \textbf{751}, 479--486 (2015)
[arXiv:1410.8822 [hep-ph]].

\bibitem{Lansberg:2015lva}
J.-P.~Lansberg and H.~S.~Shao,
Double-quarkonium production at a fixed-target experiment at the LHC (AFTER@LHC),
Nucl.\ Phys.\ B \textbf{900}, 273--294 (2015)
[arXiv:1504.06531 [hep-ph]].

\bibitem{Lansberg:2019fgm}
J.-P.~Lansberg, H.-S.~Shao, N.~Yamanaka, and Y.-J.~Zhang,
Prompt $J/\psi$-pair production at the LHC: impact of loop-induced contributions and of the colour-octet mechanism,''
Eur.\ Phys.\ J. C \textbf{79}, 1006 (2019)
[arXiv:1906.10049 [hep-ph]].

\bibitem{Sun:2023exa}
L.-P.~Sun,
$J/\psi$ pair hadroproduction at next-to-leading order in nonrelativistic-QCD at CMS,
Chin.\ Phys.\ C \textbf{47}, 093105 (2023)
[arXiv:2307.02809 [hep-ph]].

\bibitem{Sun:2023exb}
L.-P.~Sun,
$J/\psi$ pair hadroproduction at next-to-leading order in nonrelativistic-QCD at ATLAS,
Chin.\ Phys.\ C \textbf{48}, 033102 (2024).

\bibitem{He:2025kkw}
Z.-G.~He, X.-B.~Jin, and B.~A.~Kniehl,
Next-to-leading-order QCD corrections to double prompt $J/\psi$ hadroproduction,
Phys.\ Rev.\ D \textbf{111}, 094040 (2025)
[arXiv:2505.04357 [hep-ph]].

\bibitem{Li:2013csa}
Y.-J.~Li, G.-Z.~Xu, K.-Y.~Liu, and Y.-J.~Zhang,
Relativistic correction to $J/\psi$ and $\Upsilon$ pair production,
JHEP \textbf{07}, 051 (2013)
[arXiv:1303.1383 [hep-ph]].

\bibitem{He:2024ugx}
Z.-G.~He, X.-B.~Jin, and B.~A.~Kniehl,
Relativistic corrections to prompt double charmonium hadroproduction near threshold,
Phys.\ Rev.\ D \textbf{109}, 094013 (2024)
[arXiv:2402.07773 [hep-ph]].

\bibitem{He:2021oyy}
Z.-G.~He, B.~A.~Kniehl, M.~A.~Nefedov, and V.~A.~Saleev,
Double prompt $J/\psi$ production at hadron colliders,
Mod.\ Phys.\ Lett.\ A \textbf{36}, 2130018 (2021)

\bibitem{D0:2014vql}
V.~M.~Abazov \textit{et al.}\ (D0 Collaboration),
Observation and studies of double $J/\psi$ production at the Tevatron,
Phys.\ Rev.\ D \textbf{90}, 111101 (2014)
[arXiv:1406.2380 [hep-ex]].

\bibitem{He:2018hwb}
Z.-G.~He, B.~A.~Kniehl, and X.-P.~Wang,
Breakdown of Nonrelativistic QCD Factorization in Processes Involving Two Quarkonia and its Cure,
Phys.\ Rev.\ Lett.\ \textbf{121}, 172001 (2018)
[arXiv:1809.07993 [hep-ph]].

\bibitem{Nogueira:1991ex}
P.~Nogueira,
Automatic Feynman Graph Generation,
J. Comput.\ Phys.\ \textbf{105}, 279--289 (1993).

\bibitem{Vermaseren:2000nd}
J.~A.~M.~Vermaseren,
New features of FORM,
[arXiv:math-ph/0010025 [math-ph]].

\bibitem{Hahn:2004fe}
T.~Hahn,
CUBA--a library for multidimensional numerical integration,
Comput.\ Phys.\ Commun.\ \textbf{168}, 78--95 (2005)
[arXiv:hep-ph/0404043 [hep-ph]].

\bibitem{ParticleDataGroup:2024cfk}
S.~Navas \textit{et al.}\ (Particle Data Group),
Review of Particle Physics,
Phys.\ Rev.\ D \textbf{110}, 030001 (2024).

\bibitem{Kang:2011mg}
Z.-B.~Kang, J.-W.~Qiu, and G.~Sterman,
Heavy Quarkonium Production and Polarization,
Phys.\ Rev.\ Lett.\ \textbf{108}, 102002 (2012)
[arXiv:1109.1520 [hep-ph]].

\bibitem{Pumplin:2002vw}
J.~Pumplin, D.~R.~Stump, J.~Huston, H.-L.~Lai, P.~Nadolsky, and W.-K.~Tung,
New generation of parton distributions with uncertainties from global QCD analysis,
JHEP \textbf{07}, 012 (2002)
[arXiv:hep-ph/0201195 [hep-ph]].

\bibitem{Eichten:1995ch}
E.~J.~Eichten and C.~Quigg,
Quarkonium wave functions at the origin,
Phys.\ Rev.\ D \textbf{52}, 1726--1728 (1995)
[arXiv:hep-ph/9503356 [hep-ph]].

\bibitem{Ma:2010vd}
Y.-Q.~Ma, K.~Wang, and K.-T.~Chao,
QCD radiative corrections to $\chi_{cJ}$ production at hadron colliders,
Phys.\ Rev.\ D \textbf{83}, 111503(R) (2011)
[arXiv:1002.3987 [hep-ph]].

\bibitem{LHCb:2023wsl}
R.~Aaij \textit{et al.}\ (LHCb Collaboration),
Measurement of associated $J/\psi$-$\psi(2S)$ production cross-section in $pp$ collisions at $\sqrt{s}=13$~TeV,
JHEP \textbf{05}, 259 (2024)
[arXiv:2311.15921 [hep-ex]].

\bibitem{Kang:2014tta}
Z.-B.~Kang, Y.-Q.~Ma, J.-W.~Qiu, and G.~Sterman,
Heavy quarkonium production at collider energies: Factorization and evolution,
Phys.\ Rev.\ D \textbf{90}, 034006 (2014)
[arXiv:1401.0923 [hep-ph]].

\bibitem{Ma:2014svb}
Y.-Q.~Ma, J.-W.~Qiu, G.~Sterman, and H.~Zhang,
Factorized Power Expansion for High-$p_T$ Heavy Quarkonium Production,
Phys.\ Rev.\ Lett.\ \textbf{113}, 142002 (2014)
[arXiv:1407.0383 [hep-ph]].

\end{thebibliography}
\end{document}